\documentclass[iop,apj]{emulateapj}

\newcommand{\msun}{M_\odot}

\shorttitle{Ejecta-companion interaction in iPTF 13bvn}
\shortauthors{Hirai et al.}

\usepackage[plainpages=false, colorlinks=true, anchorcolor=blue, linkcolor=blue, citecolor=blue, bookmarks=false]{hyperref}

\begin{document}

\title{Possible signatures of ejecta-companion interaction in iPTF 13bvn}

\author{Ryosuke Hirai\altaffilmark{1}, and Shoichi Yamada\altaffilmark{1}}
\affil{\altaffilmark{1}Advanced Research Institute for Science and Engineering, Waseda University, 3-4-1, Okubo, Shinjuku, Tokyo 169-8555, Japan}

\begin{abstract}
We investigate the possible effects of the supernova ejecta hitting the companion star in iPTF 13bvn, focusing on the observable features when it becomes visible. iPTF 13bvn is a type Ib supernova that may become the first case that its progenitor is identified as a binary by near future observations. According to calculations by \cite{ber14}, the progenitor should have a mass $\approx 3.5\msun$ to reproduce the supernova light curve, and such compact stars could only be produced via binary evolution. This is one of the reasons that we expect the progenitor to be a binary, but it should be confirmed by observing the remaining companion after the supernova. Their evolutionary calculations suggest that the companion star will be an overluminous OB star at the moment of supernova. With a combination of hydrodynamical and evolutionary simulations, we find that the secondary star will be heated by the supernova ejecta and expand to have larger luminosities and lower surface effective temperatures. The star will look rather like a red super giant, and this should be taken into account when searching for the companion star in the supernova ejecta in future observations.

\end{abstract}


\section{Introduction}
Type Ib supernovae (SNe) are one of the hydrogen deficient subtypes of core-collapse supernovae, which are the final fates of massive ($M\gtrsim8\msun$) stars \cite[]{fil97}. Their lack of hydrogen indicates that they explode from stripped-envelope progenitors. 
Two major scenarios have been proposed to explain the removal of the hydrogen layers. One is by strong stellar winds for stars with zero age main sequence (ZAMS) masses $M_{ZAMS}\gtrsim25\msun$ \cite[]{mae81}. Such stars are called Wolf-Rayet stars, and are known that they sometimes shed their entire hydrogen envelope with the wind. The other possible scenario is mass transfer in close binaries \cite[]{pod92}. Outer layers of the more evolved star in a binary can be transferred to its companion via Roche lobe overflow, or removed dynamically in common envelope phases. Debates continue about which is the more likely scenario.

iPTF 13bvn was a recent SN of this particular type, first identified by the intermediate Palomar Transient Factory \cite[]{law09} on June 16.238 UT 2013. Its host galaxy is NGC 5806, which is at a distance $\sim21$Mpc. Pre-explosion images of \textit{Hubble Space Telescope (HST)} showed a candidate progenitor within a $2\sigma$ error of the SN site \cite[]{cao13}. It is not settled whether we were looking at the progenitor star itself or a combined flux of a binary, but nevertheless we can place strong constraints on the exploding star.

\cite{gro13} first proposed that a Wolf-Rayet star with a ZAMS mass of $31-35\msun$ could be a possible progenitor for iPTF 13bvn. This result was based on the absolute magnitude of the source in the pre-explosion image, and the strict upper limit on the radius of the progenitor ($\lesssim5R_{\odot}$) due to early detection. In their case, the pre-SN progenitor mass was $\approx10\msun$. This possibility was ruled out by observations of the later phases of the bolometric light curve \cite[]{ber14,fre14}. Detailed hydrodynamical simulations of SNe with different progenitors showed that the observed light curve cannot be reproduced by a star with $M\gtrsim8\msun$. Instead, the preferred model was a He star with $M\approx3.5\msun$ which is difficult to produce by assuming only stellar winds. The limit on the stellar radius was also extended up to $\lesssim150R_{\odot}$ by an additional set of simulations with extended thin envelopes by \cite{ber14} (hereafter B+14). Their results strongly support the binary evolution scenario, and they also showed a possible evolutionary path that can produce such a low mass progenitor star and match the pre-SN \textit{HST} observations consistently. The SN is estimated to fade below the brightness of the pre-SN primary star at about three years from explosion. A fainter secondary star may be observed shortly afterwards. If a companion star is really discovered, it will be the first case that the progenitor for a type Ib SN is confirmed to be a binary.

 According to the binary evolution calculations by B+14, the secondary star is predicted to be an overluminous OB-type star. The combined optical flux of the primary and the secondary was consistent with the optical flux from the pre-SN image. However, the SN ejecta may change the observable features of the companion after the explosion by stripping mass, or heating the star to make it swell up \cite[]{RH14}. This effect was not included in previous predictions although it may significantly alter its appearance.

 In this paper we investigate the possible effects of SN ejecta hitting the companion in iPTF 13bvn, and how it may affect the observational features. Such events have been thoroughly studied for the single-degenerate scenario of type Ia SNe \cite[]{mar00,pod03,pan13,liu13,sha13}, but not so much for massive binaries, and relatively wide binaries, which do not fit in the picture of the single-degenerate scenario. Here we first model binary systems that fit within the constraints placed by B+14 via stellar evolution calculations, and then carry out hydrodynamical simulations of the SN ejecta colliding with the companion star. To obtain the stellar structures at the time it becomes visible in the ejecta, we then perform additional stellar evolution simulations with extra heat distributed in the outskirts of the star. Parameters for the artificial heating were evaluated from the hydrodynamical simulations.

This paper is structured as follows: Our choice of model parameters and numerical method is explained in Section 2. The results of our evolution and hydrodynamical simulations are explained in Section 3, and we discuss our results in Section 4. We summarize our conclusions in Section 5.

\section{Models and Numerical Method}

B+14 revealed that the progenitor mass of iPTF 13bvn should be in the range of 3-5$\msun$, presumably $\approx3.5\msun$. By detailed binary evolution calculations, they also placed constraints on the initial configuration of the binary system that can produce such stripped envelope progenitors in this mass range. According to their estimates, the initial primary star (progenitor) mass should be $15\lesssim M^i_1\lesssim 25\msun$, and the mass ratio $0.8\lesssim M^i_2/M^i_1\lesssim 0.95$. Note that the lower bound of the mass ratio is not physically motivated, but set to avoid common envelope phases since they could not handle common envelope evolution with their stellar evolution code. These constraints overlap with the results by \cite{eld15}, where the permitted primary mass was $10\msun\lesssim M^i_1\lesssim20\msun$. The overlapping range may be most plausible. They further estimated the possible range of the final secondary mass, being $23\lesssim M^f_2\lesssim 45\msun$ if conservative mass transfer is implied. This range is expanded to $18\lesssim M^f_2\lesssim 45\msun$ if mass accretion efficiencies are taken in the range $0.5\lesssim \beta\leq1$ where $\beta$ is the fraction of transfered matter that is accreted onto the secondary star \cite[]{ben13}. The luminosity of the secondary star ended up within the constraint placed by the pre-explosion image of iPTF 13bvn.

 In this paper, we first attempt to construct a pre-SN binary model that fits within the observationally permitted range. We utilize the open source stellar evolution code \texttt{MESA} \cite[version 7184;][]{pax11,pax13} to model each star in the binary. Binary evolution is treated by evolving two spherically symmetric one dimensional stars with a mass transfer rate applied by the ``Ritter formalism'' \cite[]{rit88}. Orbital elements are also evolved consistently. We assume no rotation, no magnetic fields, and circular orbits for all models to maintain simplicity.

 It is known that even for identical initial conditions, different stellar evolution codes give different results \cite[]{jon14,suk14}. Since the code employed in B+14 \cite[a code by][]{ben13,ben03} was different from ours, we could not produce the same binary model as the example case in B+14 by taking the same initial conditions.
 Here we aim to reproduce similar pre-SN configurations as the one shown in B+14, so we take slightly different initial parameters. We also construct several other binary models for comparison. The selected binary parameters are listed in table \ref{tab_param}, along with the final binary parameters at the moment of SN. For the non-conservative mass transfer models (with $\beta=0.5$), the angular momentum loss parameter is given as $\alpha=1$ \cite[]{ben03}.
All models were calculated until itthe primary starts producing iron at the centre. The primary stars are assumed to explode within hours after iron production, and binary parameters are unchanged until explosion.

 \begin{table}
  \begin{center}
   \caption{Binary parameters of the stellar evolution calculations with \texttt{MESA}.\label{tab_param}}
   \begin{tabular}{cccccccc}
    \tableline\hline
    Model\tablenotemark{a} 
          & Age  & $M_1$& $M_2$& $R_1$& $R_2$& $P$  & $a$ \\
          & (Myr)& ($\msun$)& ($\msun$)&($R_\odot$)&($R_\odot$)&(days)& ($R_\odot$)\\
    \tableline
    a1.0  & 0    & 19.0 & 18.0 & 5.65 & 5.47 & 2.45 & 25.5 \\
          & 10.4 & 3.48 & 33.5 & 44.7 & 7.72 & 61.7 & 219  \\
    a0.5  & 0    & 19.0 & 18.0 & 5.65 & 5.47 & 2.45 & 25.5 \\
          & 10.2 & 3.72 & 25.6 & 47.8 & 10.4 & 62.0 & 203  \\
    b1.0  & 0    & 18.0 & 17.0 & 5.47 & 5.30 & 3.50 & 31.7 \\
          & 9.98 & 4.15 & 30.9 & 15.4 & 8.02 & 47.9 & 181  \\
    b0.5  & 0    & 18.0 & 17.0 & 5.47 & 5.30 & 3.50 & 31.7 \\
          & 9.94 & 4.23 & 23.9 & 10.6 & 7.76 & 54.4 & 184  \\
    c1.0  & 0    & 15.0 & 14.0 & 4.93 & 4.74 & 3.30 & 28.7 \\
          & 12.5 & 3.02 & 26.0 & 31.8 & 7.01 & 63.4 & 206  \\
    c0.5  & 0    & 15.0 & 14.0 & 4.93 & 4.74 & 3.30 & 28.7 \\
          & 12.5 & 3.06 & 20.0 & 30.4 & 6.57 & 73.3 & 210  \\
    \tableline
   \end{tabular}
   \tablenotetext{1}{Numbers in the model names indicate values of the accretion efficiency parameter $\beta$ \cite[]{ben13} applied in our evolution calculations.}
   \tablecomments{All models were assumed to have circular orbits.}
  \end{center}
 \end{table}

 To the pre-SN binary model, we then apply the same method as in \cite{RH14} to simulate the SN ejecta hitting the companion. A two step strategy is employed; 1) artificially explode the primary star on a one dimensional spherical grid, 2) simulate the effect of SN ejecta hitting the companion star on a two dimensional axisymmetrical grid. The hydrodynamical code ``yamazakura'' \cite[]{saw13} is used for all hydrodynamical simulations throughout this paper. It is a mesh-based central scheme code with an iterative Poisson solver to solve self-gravity. An ideal-gas equation of state was applied with an adiabatic index $\gamma=5/3$.

For the first step, $E_{bind}+E_{exp}$ of internal energy is added to the inner few meshes of the primary star model placed on a one dimensional computational domain to initiate an explosion like a thermal bomb \cite[]{yng07}, where $E_{bind}$ is the gravitational binding energy of the star and $E_{exp}$ is the explosion energy. In order to leave a residual neutron star after the explosion, we cut the central $1.4\msun$ of the star out of the computational region, and set a reflective inner boundary condition. In this way, all the energy applied will be placed just above the neutron star and it will initiate a shock wave that propagates outwards through and out of the star. The explosion energy was taken as $E_{exp}=8\times10^{50}$erg from the estimate by B+14,  which was obtained to fit the peak of the bolometric light curve and photospheric velocity evolution \cite[See also][for discussions on the explosion energy]{sri14}. A dilute circumstellar matter is placed around the star due to numerical reasons, with a low density so as not to disturb the propagation of the SN ejecta. An outgoing outer boundary condition is employed so the ejecta can flow out freely. Ejecta profiles are sampled at a point far from the stellar surface (50 times the stellar radius), and it is checked that it follows a homologous expansion by comparing the time evolution at different points.

In the second step we place the secondary star model onto the origin of an axisymmetric $600(r)\times180(\theta)$ spherical grid. The axis is taken along the line connecting the centre of both stars. Axisymmetry is justified due to the short timescale of SN ejecta flowing past the star ($\sim1$ day) compared to the orbital period ($\sim60$ days). Unlike \cite{RH14}, we do not leave out the central portion of the star since the companion star model is not so centrally concentrated. The density scale height is resolved with at least 10 meshes, with a total $\sim150$ radial gridpoints inside the star. Mesh sizes are then increased monotonically outside the star, until the radius of the computational region reaches up to $\sim90\%$ of the binary separation. Yet again we place a low density atmosphere around the star that is dilute enough so that it has negligible mass. Data from the first step is used to place the SN ejecta close to the companion as an initial condition. For the outer boundary condition, the ejecta data is extrapolated as a Dirichlet boundary on the side facing the primary, whereas a free boundary is applied on the opposite side.

Each mesh is marked as bound or unbound using the ``Bernoulli criterion'', in which matter is bound when 
\begin{equation}
 \frac{1}{2}v^2+\epsilon+\frac{p}{\rho}+\phi<0
\end{equation}
where $\epsilon$ is the specific internal energy and $p$ is pressure, $\rho$ is density and $\phi$ is the gravitational potential. We integrate over the bound region to evaluate the mass and position of the centre of mass of the remaining star. In order to see whether matter from the ejecta can mix with the original stellar matter, we also placed tracer particles that just follows the fluid motion. Each particle carries information of its mass and origin (stellar or ejecta), and is evaluated whether it is bound or not at each step.

\section{Results}

\subsection{Stellar evolution}
\begin{figure}
 \plotone{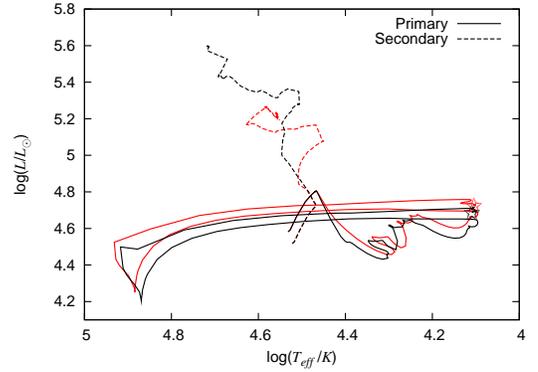}
 \caption{Evolutionary tracks of the stellar components in models a1.0 (black lines) and a0.5 (red lines). Solid lines show results of the primary stars, and dashed lines the secondary stars. The star signs label the position of the primary star at the point of SN.\label{HRDa1.0}}
\end{figure}

In this section we show results of our stellar evolution calculations. Figure \ref{HRDa1.0} shows the evolutionary track on the HR diagram of both stellar components for models a1.0 and a0.5. They roughly resemble the track of the model presented in B+14. The overall evolution of the primary does not change much for different values of $\beta$, nor does the period of the system. On the other hand, the secondary mass and luminosity depends strongly on the value of $\beta$. However, the optical flux is dominated by the primary\footnote{Even though the secondary star has a higher luminosity, the primary has a stronger optical flux due to the lower temperature.}, so it is difficult to constrain the secondary luminosity from the HST pre-SN image. All other models followed similar evolutionary tracks on the HR diagram.

 Every primary star model was deficient of hydrogen at the point of SN, which is a necessary condition to produce a type Ib SN. Final luminosities of the primary stars range $4.5\lesssim L_1^f/L_\odot\lesssim4.9$, which roughly fits in the allowed range to match the HST pre-SN flux. This is also the case for the companions, with luminosities $4.9\lesssim L_2^f/L_\odot\lesssim5.6$. From the above facts, we assume that all of our models can be progenitor candidates of iPTF 13bvn. Since the progenitor mass is most likely $\approx3.5\msun$, we will take model a1.0 to be our reference model. The b series have primary masses close to the lower limit of the observational constraint and the c series close to the upper limit.

\begin{figure}
 \plotone{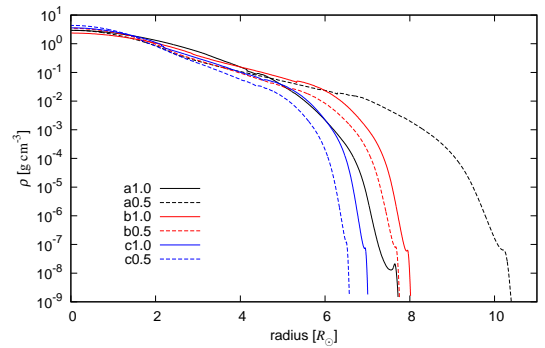}
 \caption{Density profiles of the companion star at the point of SN of the primary. Model names correspond to the names in table \ref{tab_param}.\label{comp}}
\end{figure}

Density-radius profiles for each companion star model are compared in Figure \ref{comp}. All models have similar structures, with steep density gradients near the surface and slightly different radii according to their accretion histories.

 Obviously, these are not the only models which can reproduce the observational characteristics of iPTF 13bvn. Common envelope phases may have occured to produce the compact progenitor, or the orbit may have been eccentric to induce periodic mass loss. Single star models cannot be excluded if we consider stellar rotation \cite[]{eld15}. Our stellar evolution calculations do not include these effects, and it is out of the scope of this paper to consider every possible scenario.

\subsection{Supernova}
\begin{figure*}
 \plotone{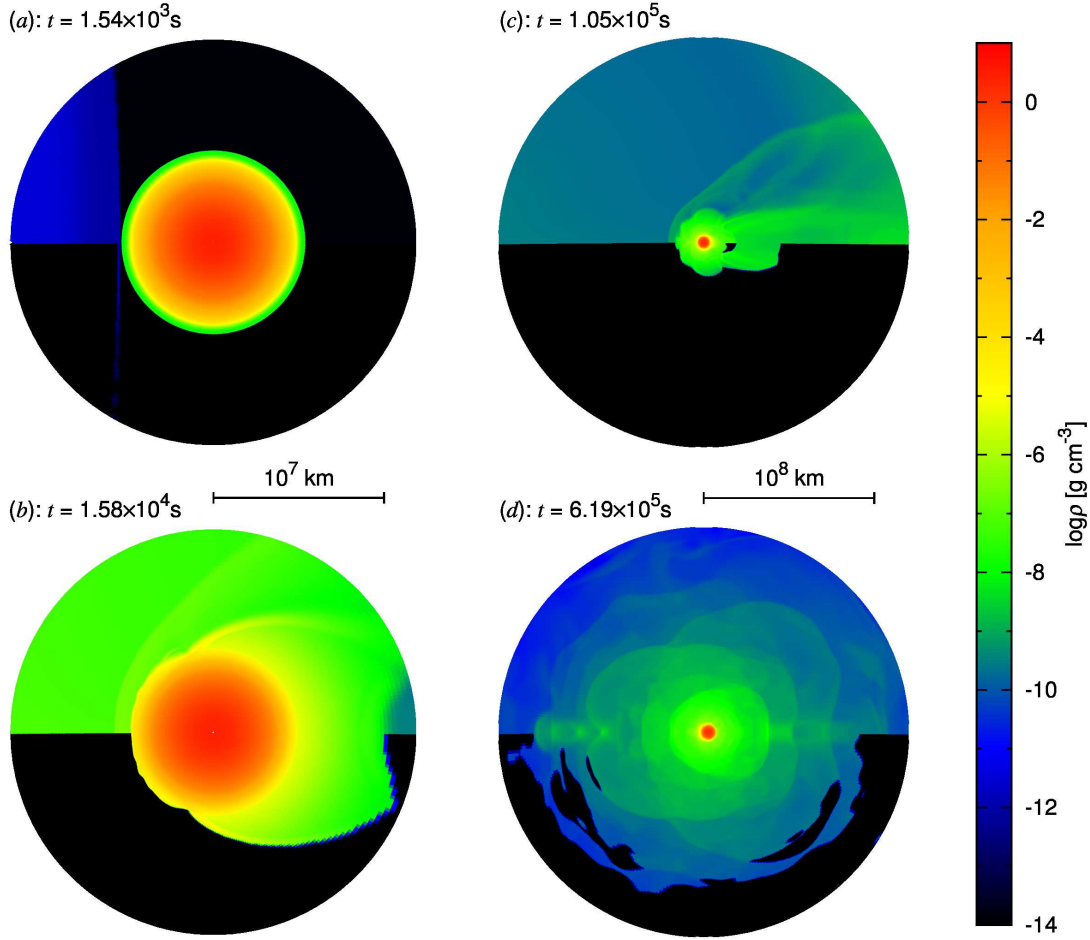}
 \caption{Density plots from the hydrodynamical simulations of the collision of SN ejecta and the companion star. Each snapshot is labelled with the time ellapsed since SN.  Only the bound matter is coloured in the lower half of each panel. The SN ejecta is flowing in from the left side of each panel. The left panels have a radius of $1.3\times10^7$km, and right panels $1.3\times10^8$km.\label{snapshot1}}
\end{figure*}

\begin{figure*}
 \plotone{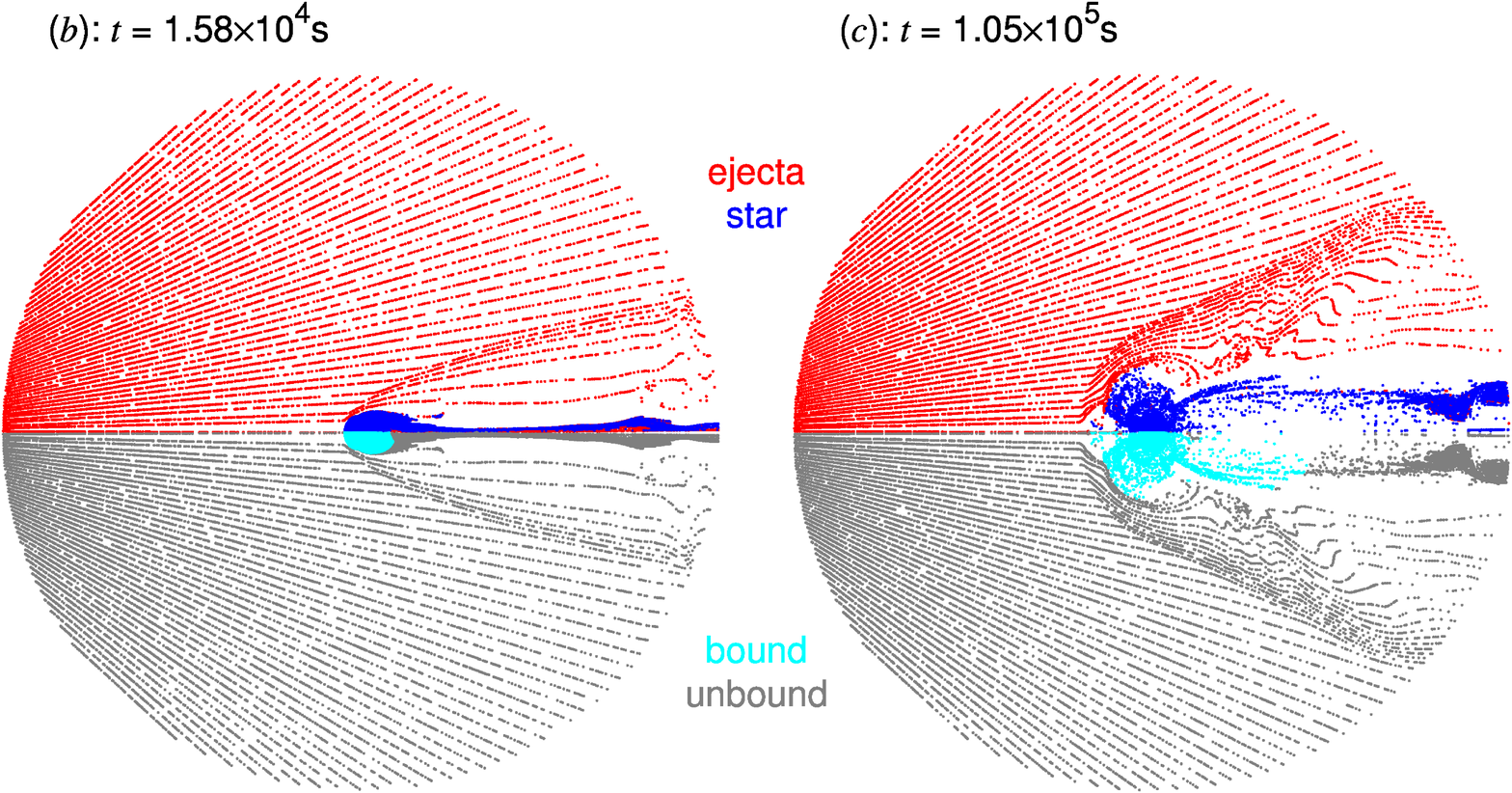}
 \caption{Distributions of tracer particles at various times. In the upper half of each panel, each particle is coloured red or blue if they originate from the ejecta or star respectively. In the lower half, particles are coloured light blue or grey depending on whethere they are bound or not. Particles that are red in the upper half and light blue in the lower half are ejecta matter that has mixed into the stellar matter. The radii of the circles are $1.15\times10^8$km.\label{snapshotptc1}}
\end{figure*}

Figure \ref{snapshot1} shows snapshots of our hydrodynamical simulations of the collision of SN ejecta and the companion star in our reference model. Note that the scales are different in the left and right panels. The whole computational region is displayed in the right panels whereas we show a close-up view in the left panels. Panel \textit{a} shows our initial condition, where the ejecta is just about to reach the companion surface. As the ejecta hits the surface of the companion, a forward shock is driven into the star, and a reverse bow shock is formed in front of the star which can be seen as density discontinuities in Panel \textit{b}. The forward shock heats the outskirts of the star as it proceeds, but the propagation soon stops at a certain point because of the pressure gradient in the star. After the bulk of the ejecta flows past, the bow shock and the heated matter starts to expand outwards as can be seen in the lower half of Panel \textit{c}. We can check whether the stellar matter is expanding, or it is just the bound region that is expanding by looking at the tracer particle distributions. Figure \ref{snapshotptc1} shows the distributions of the tracer particles at times corresponding to Panels \textit{b} and \textit{c} in Figure \ref{snapshot1}. The upper half is colour coded according to the origin of each particle, red for ejecta matter that comes from the primary, and blue for secondary star matter. The lower half is colour coded whether it is bound (light blue) or not (grey). We can see that some of the particles from the stellar surface is stripped away along the axis, but the mass is very small. If there are any particles that are red in the upper half and light blue in the lower half, it means that ejecta matter has accreted onto the companion. However, most of the light blue particles in the lower half are blue in the upper half, which indicates that the enlargement of the bound region is not due to ejecta matter being accreted to the stellar surface, but the shocked stellar matter expanding because of the heat. A small amount of ejecta matter is mixed into the stellar matter, but their masses are extremely small. The bound region keeps on expanding in a spherical manner, eventually reaching the outer boundary of our simulation. Panel \textit{d} in Figure \ref{snapshot1} shows a snapshot at these later stages, where the bound region is almost spherical.

\begin{figure}
 \plotone{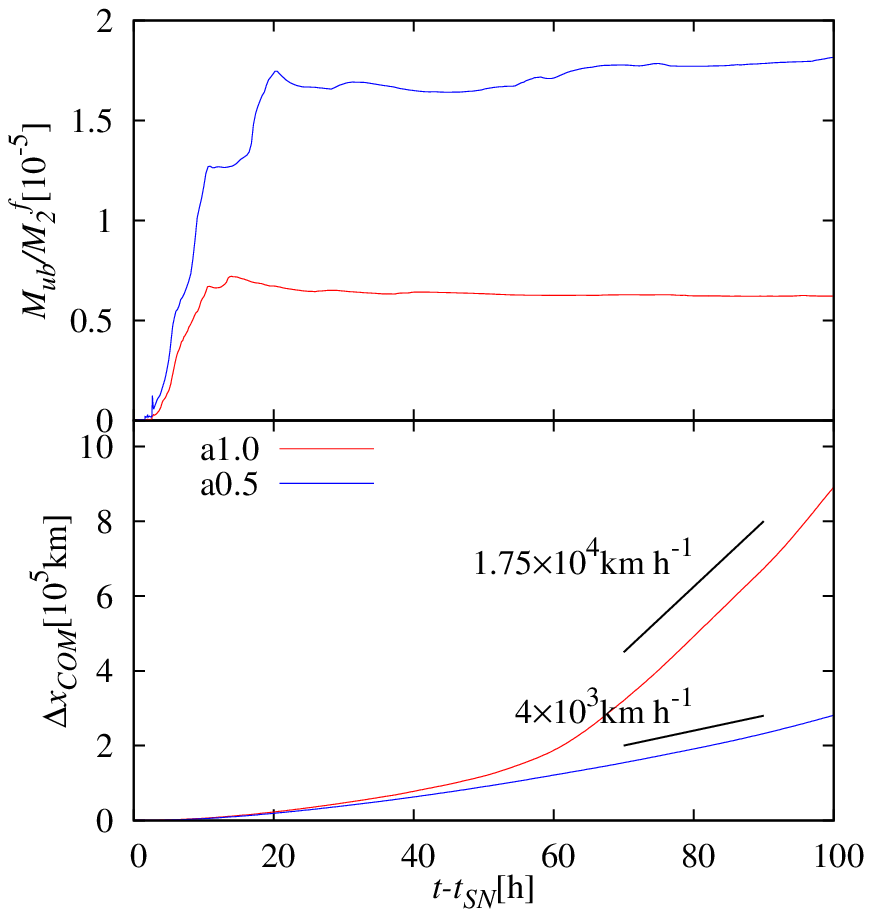}
 \caption{Time evolution of the fraction of unbound mass (upper panel) and the displacement of the centre of mass (lower panel). Red lines show results for model a1.0, and blue lines for a0.5. Black slopes in the lower panel show the approximate velocities of the centre of mass.\label{Mub_vkick}}
\end{figure}

In Figure \ref{Mub_vkick} we show the amount of unbound matter ($M_{ub}$), and the displacement of the centre of mass ($\Delta x_{COM}$) as a function of time since SN. As we can see from the upper panel, hardly any matter was stripped off nor accreted onto the star. The absolute values are comparable to the errors arising from the limitation in numerical resolution. This is consistent with previous works \cite[]{RH14,mar00,pan10}, where the unbound mass depends on the orbital separation as $\propto a^{m_{ub}}$, where $a$ is the separation. The exponent $m_{ub}$ depends on the stellar structure, but simulations suggest values $-4.5\lesssim m_{ub}\lesssim -3$. Since our models all have relatively wide separations, it is natural that the amount of stripped mass was so small. The lower panel shows the motion of the centre of mass, which reaches a constant velocity in the direction opposite to the exploding star in about a few days. According to simple estimates \cite[]{che74,whe75}, this so-called ``rocket effect'' or ``kick'' velocity\footnote{The kick velocity mentioned here is different from the natal kick imparted to the central neutron star in core collapse SNe.} should be very small in our present model, due to the wide separation with respect to the stellar radius. In our simulation, the kick velocity reached up to $1.75\times10^4$km h$^{-1}$, which was $\sim30\%$ of the orbital velodity. This relatively large kick velocity may help destroy the binary or at least alter the eccentricity of the post-SN system. For model a0.5, the kick velocity was $\sim5\%$ of the orbital velocity, which is almost negligible. All other models showed qualitatively similar results.

 In Table \ref{tab_mubvkick} we list the amount of unbound mass and the kick velocity along with the final temperature and luminosity which is explained in Section \ref{sec_dis}. It is clear that the unbound masses were negligible in every model. The kick velocity was also negligible compared to the orbital velocities. There seems to be no strong correlation between the binary parameters to the unbound mass or kick velocity. This is because the stellar structure and SN ejecta profiles are different for each model, and the absolute values are all comparable to the numerical resolution.

\begin{table}[h]
 \begin{center}
  \caption{Final states of the secondary star.\label{tab_mubvkick}}
  \begin{tabular}{cccccc}
   \tableline \hline
   Model\tablenotemark{a} & $M_{ub}$ & $v_{kick}$ & $v_{orb}$\tablenotemark{b} & log$T_{eff}\tablenotemark{c}$ & log$L\tablenotemark{c}$ \\
        & ($\msun$) &(km h$^{-1}$)& (km h$^{-1}$)& (K)&($L_\odot$)\\\tableline
   a1.0 & $2.1\times10^{-4}$ &$1.8\times10^4$& $6.1\times10^4$ & 3.82 & 5.51 \\
   a0.5 & $4.7\times10^{-4}$ &$4.3\times10^3$& $7.6\times10^4$ & 3.64 & 5.65 \\
   b1.0 & $1.1\times10^{-4}$ &$5.6\times10^3$& $8.2\times10^4$ & 4.01 & 5.85 \\
   b0.5 & $3.1\times10^{-4}$ &$9.1\times10^3$& $9.3\times10^4$ & 3.66 & 5.58 \\
   c1.0 & $0.9\times10^{-4}$ &$3.3\times10^3$& $6.1\times10^4$ & 3.61 & 5.57 \\
   c0.5 & $2.2\times10^{-4}$ &$5.7\times10^3$& $6.9\times10^4$ & 4.22 & 4.98 \\
\tableline
  \end{tabular}
  \tablenotetext{1}{Names of models correspond to those in Table \ref{tab_param}.}
  \tablenotetext{2}{The orbital velocity prior to SN.}
  \tablenotetext{3}{$T_{eff}$ and $L$ are values estimated three years after SN.}
 \end{center}
\end{table}

\section{Discussions}

The dynamical effects of the ejecta hitting the companion star, such as mass stripping and momentum transfer seem to be limited. But the expansion of the outer layers of the star may become important when we observe the companion after the SN ejecta becomes faint enough. Such expansions have also been found in previous simulations \cite[]{RH14,liu13}. Here we explore the observable features of the companion star when it becomes visible.

\subsection{Reddening due to SN Heating}\label{sec_dis}
\begin{figure}
 \plotone{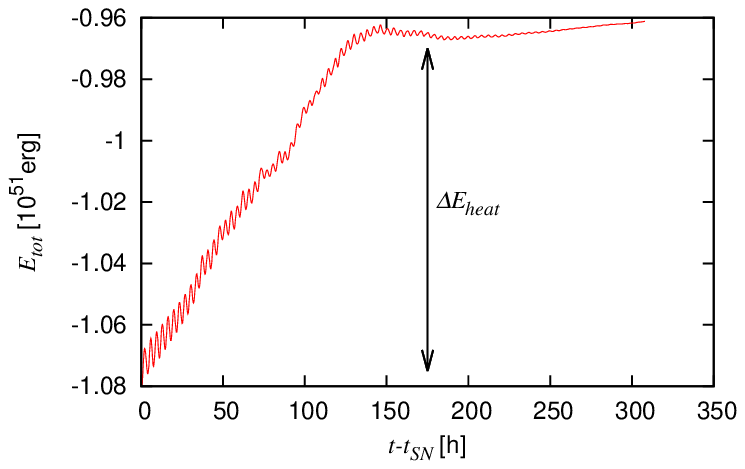}
 \caption{Evolution of the total energy of the star. Total energy is evaluated by Eq.\ref{etot_eq} at every time step. The difference between the initial and final energies are considered to be the heat imparted by the SN ejecta $\Delta E_{heat}$.\label{Etot}}
\end{figure}

In Figure \ref{Etot} we illustrate the evolution of the total energy of the star, which we calculate as
\begin{eqnarray}
 E_{tot}=\int_{V_{b}}(\phi+\epsilon+v^2)\rho dV\label{etot_eq}
\end{eqnarray}
where $V_b$ is the bound region.
 The SN ejecta imparts energy to the star via shock heating, and the star adjusts its structure to restore mechanical equilibrium. After a few days, the total energy of the star reached a constant value. We consider that the heating process has finished here, and assume that the difference between the initial and final energy is the amount of energy injected to the star by the SN ejecta.

In our hydrodynamical simulations, the expansion of the star did not cease even at the end of our simulations (a few days) where the bound region reaches the outer boundary. The numerical cost is extremely expensive to wait for the star to recover equilibrium. Hence we use the stellar evolution code instead of the hydrodynamical code to estimate the state of the star for 100 years after SN. We believe that this is justified because the expansion was almost spherical in the hydrodynamical simulations. This is a similar procedure to previous studies that predicted the post-SN appearance of the companion in type Ia SN progenitors in the single degenerate scenario \cite[]{pod03,pan12,sha13}.

\begin{figure}
 \plotone{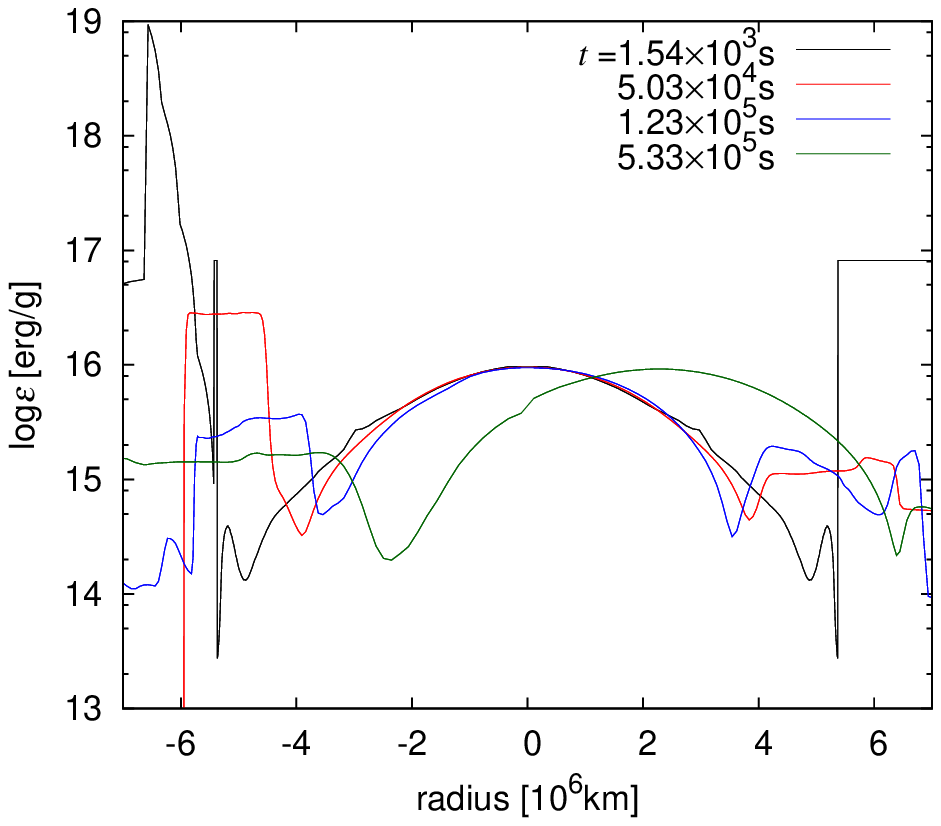}
 \caption{Time evolution of the distribution of specific internal energy along the axis. The direction of the axis is taken away from the exploding star.\label{shock}}
\end{figure}

Starting from the companion model at the pre-SN stage of the \textit{primary}, we artificially input energy into the outer layers as
\begin{eqnarray}
 \dot{\epsilon}_{heat}(m)=\frac{\Delta E_{heat}}{t_{heat}\sqrt{2\pi\sigma^2}}\textrm{exp}\left(-\frac{(m-\mu)^2}{2\sigma^2}\right)\label{heateq}
\end{eqnarray}
where $\epsilon_{heat}$ is the artificial specific internal energy injection rate as a function of mass coordinate, $\Delta E_{heat}$ is the total energy input, $t_{heat}$ is the duration of heating, and $\sigma=(M_2^f-m(r_{in}))/6$, $\mu=(M_2^f+m(r_{in}))/2$. $r_{in}$ is the radial coordinate at the inner edge of the heating layer, where the shock could not proceed further inwards in the hydrodynamical simulations. Figure \ref{shock} illustrates the time evolution of the distribution of specific internal energy along the axis. The black line shows the initial condition, where we can see the ejecta approaching the stellar surface ($\sim5.3\times10^6$km) from the left. The forward and reverse shocks can be seen as a steep wall in the red line, and weakens its strength as it proceeds into the star which can be seen in the blue line. The forward shock propagates up to $\sim3\times10^6$km from the centre at most. We can see from the green line, the internal energy distribution near the end of our simulation, that the stellar matter interior to this point is almost unaffected by the shock. We therefore take $r_{in}=3\times10^6$km as the inner edge of the heating layer. $\Delta E_{heat}$ and $t_{heat}$ are also set to the values taken from the hydrodynamical simulations ($\Delta E_{heat}=1\times10^{50}$erg, $t_{heat}=100$hrs). Our strategy of extra energy deposition is similar to \cite{sha13} and \cite{pod03}. The main difference is that we ignore the effects of mass stripping, since the amount of stripped mass in our hydrodynamical simulations were negligible (cf. table \ref{tab_mubvkick}). Due to this assumption, all the energy from the SN ejecta is used to puff up the star, not to ablate away some surface matter.

\begin{figure}
 \plotone{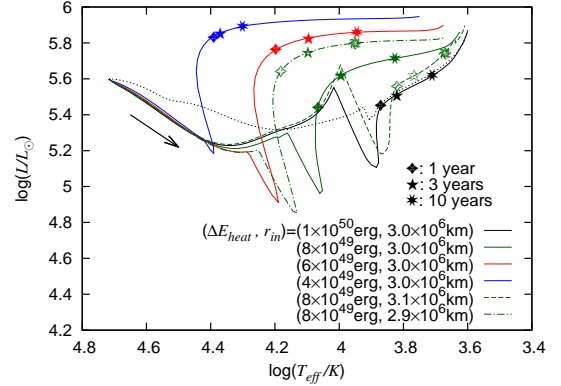}
 \caption{Evolution of the secondary star after SN of the primary for model a1.0. The original position is the left end of the lines, and the right ends are 100 years after SN. Star symbols mark the location at 1, 3 and 10 years since SN. Different colours show results with different $\Delta E_{heat}$, and the dashed and dot-dashed lines show results with different $r_{in}$. The black dotted line shows the evolution of the star with the same heating parameters as the black solid line, but with $t_{heat}=0.2$ years.\label{HRDheat}}
\end{figure}

The shock heating of the outer layers by the SN ejecta modifies the stellar structure, pushing out the outer layers like it did in the hydrodynamical simulations. In Figure \ref{HRDheat} we show the evolutionary track of the artificially heated companion star on the HR diagram for 100 years after SN. Focusing on the fiducial model (black line), it can be seen that the temperature decreases almost monotonically. During the heating phase, temperature and luminosity both drops but the luminosity eventually starts to increase due to the growing expansion speed. As soon as we switch off the artificial heating, the temperature and luminosity shows an almost discrete drop because of the disappearance of heat\footnote{The artificial heat is applied all the way up to the surface, contributing to the surface temperature.}. In the following few dynamical times, the radius and luminosity increases to restore hydrostatic equilibrium. We do not expect these large fluctuations in the early phase to be realistic, since it is highly affected by our methodology. Thus we only pay attention on the evolution after several dynamical times. After a year, the effective temperature will drop by an order of magnitude, and the luminosity slightly increases. It was roughly estimated in B+14 that three years is the time it takes for the SN light curve to decline below the brightness of the original progenitor. The time it takes for the companion to become visible depends on the declination of the SN light curve and the band used for the observation, but we expect that it is not so far from three years, maybe slightly earlier. At that time, it will be somewhere around the middle star sign in Figure \ref{HRDheat}, somewhat like a red supergiant. Even after it becomes visible, we expect the star to continue increasing its luminosity for a few decades.

Although we have carefully chosen our parameters $r_{in}$, $\Delta E_{heat}$, and $t_{heat}$ to match our hydrodynamical simulations, there are still some uncertainties left. Results for the same simulation with different parameters are also shown in Figure \ref{HRDheat}. The final temperature strongly depends on $r_{in}$, where larger values of $r_{in}$ led to lower temperatures and lower luminosity. This is because if $r_{in}$ is larger, the extra energy is injected into a much smaller volume, so it needs to expand more to retain equilibirum. Final states are also sensitive to the value of $\Delta E_{heat}$, which can be seen in Figure \ref{HRDheat} where results for different $\Delta E_{heat}$ are plotted with different colours. Larger values of $\Delta E_{heat}$ give lower temperatures and higher luminosities. We have also conducted the same simulation with different $t_{heat}$ (black dotted line in Figure \ref{HRDheat}), but the differences were limited. As explained earlier, the early evolution is largely affected by our methodology, particularly on $t_{heat}$, but the differences become indistinguishable within a few dynamical times after the heating phase. We are therefore sure that the long term evolution will not depend on our choice of $t_{heat}$, and any earlier evolution (which is not observable anyway) should be studied by hydrodynamical simulations since it is highly unspherical. Regardless of the uncertainties in our assumptions, the secondary star will appear as a red star when it becomes visible with temperatures lower by a factor of $\sim$2-10, off the main sequence. The parameter dependence of the luminosity is weak, just slightly increasing in some cases by less than a factor of $\sim$2.

Estimates of the final temperature and luminosity for all models are listed in Table \ref{tab_mubvkick}. All other models have considerably lower temperatures compared to their original states. The expansion depends strongly on the stellar structure so it may be possible to distinguish among models from observations.

\subsection{Other Consequences}
The expansion of the star exceeded the original binary separation in some of our models, which means that it may engulf the newly born neutron star. For such cases, the pulsar wind emitted by the neutron star may heat or strip the loosely bound envelope and distort the shape. If the neutron star becomes embedded in the expanded envelope, it might initiate a common envelope phase, possibly leading to much tighter orbits or the formation of Thorne-\.Zytkow objects \cite[]{tho77}. It will be difficult to confirm whether the system has entered a common envelope phase by observations, but if the red super giant-like companion is found and no neutron star is detected for the following couple of years, it may be the first ever evidence of an ongoing common envelope phase.

\section{Conclusion}
iPTF 13bvn is a candidate of the first case that the progenitor for a type Ib SN is confirmed to be a binary. There are several possible evolutionary paths that can both reproduce the SN light curve and the pre-SN photometry of iPTF 13bvn, all containing a compact ($\lesssim 5\msun$) He star primary and an overluminous OB secondary star. We have studied the effects of SN ejecta colliding with this secondary OB star, and we found that the surface of the star may swell up with the heat, becoming more red and luminous when it becomes visible in the SN remnant. The change in temperature and luminosity strongly depends on the parameters $\Delta E_{heat}$ and $r_{in}$ that we used in the post-SN evolution calculations. But in any case the secondary will be found off the main sequence, with lower effective temperatures and a slight increase in luminosity which also depends on the time we detect it. Effects of mass stripping and kicks are limited, and will not affect the further evolution of the binary.

The radius of the secondary star becomes so large that it may engulf the primary-produced neutron star. In that case we might be able to observe the first ever ongoing common envelope phase as an absence of the neutron star in the vicinity.

We note that other evolutionary paths are possible if we include stellar rotation, common envelope phases or eccentric orbits. Theories that can deal with all of these cases have not been established yet and it is much out of the scope of this paper \cite[See][for a review on common envelopes]{iva13}. If we concern these possibilities, there is a much wider range of possible progenitors so further studies are certainly warranted.

\acknowledgements
This work was supported by the Grants-in-Aid for the Scientific Research (A) (NoS. 24244036), MEXT Grant-in-Aid for Scientific Research on Innovative Areas ``New Developments in Astrophysics Through Multi-Messenger Observations of Gravitational Wave Sources'' (Grant Number A05 24103006)

\clearpage

\end{document}